\documentclass[fleqn,twoside]{article}%
\usepackage{amsmath}
\usepackage{amsfonts}
\usepackage{amssymb}
\usepackage{graphicx}
\usepackage{cite}
\def\be{\begin{equation}}

\def\ee{\end{equation}}
\def\bes{\begin{equation}\begin{split}&}
\def\es{\end{split}}
\def\bi{\bibitem}

\begin{document}
\title{New conserved tensors and Brans-Dicke type field equation, using integrability condition.}

\author{Abhik Kumar Sanyal\footnote{E-mail: sanyal\_ ak@yahoo.com}~, Bijan Modak\footnote{E-mail: bijanmodak@yahoo.co.in}~, Manas Chakrabortty\footnote{E-mail: manas.chakrabortty001@gmail.com}}

\maketitle
\noindent
\begin{center}
\noindent
$^*$ Dept. of Physics, Jangipur College, Murshidabad, West Bengal, India - 742213. \\
$^\dag$ Dept. of Physics, University of Kalyani, Nadia, West Bengal, India - 741235.\\
$^\ddag$ Dept. of Physics, Bankura University, Bankura, West Bengal, India - 722155. \\
\end{center}

\noindent
\abstract{We explore some new off-shell and on-shell conserved quantities for a scalar field in Minkowski space, using integrability condition. The off-shell conserved tensors are related to the kinematics of the field, while a linear combination of the off-shell and the on-shell conserved tensors ends up with the energy-momentum tensor for the scalar field. In the curved background, using Ricci and Bianchi identities, scalar-tensor theory of gravity, that resembles somewhat with Brans-Dicke type field equations, emerges without requiring the principle of equivalence. Further, starting from the curvature scalar and using these identities, the field equations for modified gravity (Einstein-Hilbert action in the presence of higher-order terms) follows.}\\

\noindent
keywords: Field theory; Ricci Identity; Conserved currents.\\

\maketitle
\flushbottom

\section{Introduction}

The study of the scalar field is as old as the classical field theory. However, only recently `phi' has been detected in the sky, in the form of Higgs boson \cite{Higgs1, Higgs2, Higgs3, Higgs4, Higgs5, Higgs6, Higgs7, Higgs8}, and interest in the study of scalar field is raised. The field equations apparently admit energy-momentum tensor to be the only conserved quantity, which is therefore on-shell. Nevertheless, additional on-shell symmetries may also sometimes be found under Noether symmetry analysis. In fact, there is a general belief that all the on-shell symmetries are Noether symmetries, which carry a Noether conserved current. Nonetheless, there are examples in which, some general symmetries together with their associated conserved currents have been explored by manipulating the field equations of non-minimally coupled scalar-tensor theory of gravity, even in the presence of higher-order curvature invariant terms \cite{NO1, NO2, NO3, NO4, NO5, NO6}. It has also been proven in view of the vector field (generator), that such aforesaid conserved currents lead to inconsistency in the Noether equations, and therefore these symmetries are non-Noether \cite{NO3}. It therefore appears that additional symmetries may be found following different techniques, which are altogether different from Noether symmetry.\\

The existence of the solution of the scalar field equation, obtained under appropriate boundary conditions, may be stated in a way such that the field and its gradient are integrable in the flat space-time. These integrability conditions are not independent, rather are in-built in the theory, and therefore these are not required to find solutions. However, here we show that additional conserved tensors may be explored as a consequence of these conditions, both in the flat as well as in the curved space-time. The first type of these are off-shell and therefore independent of the field equations. The other types are on-shell and therefore are the consequences of the field equations. We relate the non-dynamical, or the geometric conserved tensors with the kinematical quantities of the scalar field, and a linear combination of the dynamical and non-dynamical ones with the stress-energy tensor of the field. Even though it is well known that one can add quantities that are trivially conserved in the energy-momentum tensor of a theory, the derivation of such tensors through purely geometric arguments is certainly useful \cite{blas:np}.\\

In view of Mach Principle the local inertial frame is determined by the average motion of distant objects. The gravitational coupling may be scale dependent, and as a consequence, the concept of inertia and the equivalence principle have to be designed or modified in a manner such that the gravitational constant is no longer a constant, rather dependent on the scalar field in the generalised action containing Einstein-Hilbert term being non-minimally coupled with the field. A special type of non-minimally coupled scalar-tensor theory of gravity is the Brans-Dicke theory of gravity. It is possible to reduce both the generalised scalar tensor theory of gravity and $F(R)$ theory of gravity under conformal transformation, to the usual Brans-Dicke form in Jordan frame with massive potential. Noether symmetry may thereafter be applied to explore a form of the potential and hence solutions. For example, equivalence between a generalised (with arbitrary coupling parameter $f(\phi)$) scalar-tensor theory and Brans-Dicke theory has been established and cosmological solutions have been explored in the presence of a barotropic fluid \cite{16}. Starting from a generalised scalar-tensor theory of gravity, equivalence between the Jordan and the Einstein frames is established under conformal transformation. Such equivalence is utilised to find exact solutions in which scalar fields are coupled with geometry \cite{17}. Again in reference \cite{18}, a nonstandard generalized theory of gravity has been translated to the Jordan frame under conformal transformation and the fact that Noether symmetries are conformally preserved in general, has been established. Further, viable cosmological solution of $F(R)$ theory of gravity in the presence of barotropic fluid has been found in view of Noether symmetry, and the link of these solutions with scalar-tensor theory of gravity has been established \cite{19}. In the present work instead of considering conformal transformation and Noether symmetry we, on the contrary, consider Klein Gordon equation for the scalar field $\phi$, together with Bianchi identity and Ricci identity for a vector constructed from a scalar function, to formulate Brans-Dicke type field equation, assuming $\phi$ to be a slowly varying function.\\

In the following section, we find off-shell and on-shell symmetries of the Klein-Gordon equation in flat (Minkowski) space-time, and also attempt to give appropriate physical interpretation of such conserved tensors. In section 3 and section 4, the same has been performed both in the flat as well as in the curved space-time respectively, with complex scalar field. In section 5 we present a conserved tensor with a real scalar field $f(\phi)$ and explore its similarity with the generalized Brans-Dicke field equation. Next, in section 6 we present a class of conserved tensors starting from a vector, built from the $n\mathrm{^{th}}$ power of the Ricci scalar, which has been generalized for $F(R)$ thereafter. Section 7 concludes our findings.

\section{Real Scalar field in flat (Minkowski) space-time:}

Consider a Lagrangian density for a real scalar field in flat space-time, $\eta_{\alpha\beta} = \mathrm{diag}(1,-1,-1,-1)$ as,

\be \label{2.1}\mathcal{L} = \frac{1}{2} \phi_{,\alpha}\phi^{,\alpha} - V(\phi); ~~\mathrm{where},~~\phi_{,\alpha} = \frac{\partial\phi}{\partial x^\alpha}, \ee
which leads to the Klein-Gordon equation,

\be \label{2.2}\Box\phi + \frac{\partial V}{\partial \phi} = 0; ~~\mathrm{where},~~\Box{\phi} = \phi^{,a}{_{,a}}.\ee
To explore the conserved tensors, we use the Ricci identity for an arbitrary vector $B^a$ in flat space-time as,

\be \label{2.3} B^a{_{,c,b}} -B^a{_{,b,c}} = 0,\ee
which is the integrability condition of a vector field $B^a$, in disguise.

\subsection{Non-dynamical (off-shell) conserved tensors:}

Being a vector, the gradient of the scalar field must satisfy the above form of integrability condition \eqref{2.3} and hence one can write,

\be\label{2.4} \phi^{,a}{_{,b,a}} - \phi^{,a}{_{,a,b}} =0,\ee
where, contraction has been performed. Now rearranging the indices, we can express the above equation as

\be \label{2.5}\phi^{,a}{_{,b,a}} - \phi^{,c}{_{,c,a}}\delta ^a{_b} = 0,\ee
which may finally be expressed as,

\be \label{2.6} T_1{^a{_{b,a}}} = 0,~~\mathrm{where},~~T_1{^a{_b}} = \phi^{,a}{_{,b}} - \phi^{,c}{_{,c}}\delta ^a{_b}.\ee
The conservation law is off-shell and is true for any scalar function. $T_{1ab}$ is obtained from a geometric identity, hence it is a geometric conserved tensor. The tensor $T_{1ab}$ in (6) is symmetric as $\phi_{,a,b}=\phi_{,b,a}$. However in non-symmetric $T_{1ab}$ one can add asymmetric tensor field to $T_{1ab}$ in order to obtain a symmetric conserved tensor. The new improved tensor $\tau_{1ab}$ following \cite{blas:np} is $\tau_{1ab} =T_{1ab}+ {S_{cab}}^{,c}$, where $S_{cab} = -S_{acb}$ is the Belinfante-Rosenfeld tensor. The tensor $T_{1ab}$ may be generated also from $S^{cab}= \phi^{,a}\eta^{cb}-\phi^{,c}\eta^{ab}$ \cite{blas:np}, since $T_{1ab}= {S^{cab}}_{,c}$ is true trivially.\\

\noindent
To find even more such geometrical (off-shell) conserved tensors, let us start from the square of the scalar field $\phi^2$, say for example. The integrability condition of the vector $\phi^2_{,a}$, then reads as,

\be \label{2.7}(\phi^2)^{,a}{_{,b,a}} - (\phi^2)^{,a}{_{,a,b}} = 0.\ee
Equation \eqref{2.7} may finally be expressed under the change of dummy index appropriately, as

\be \label{2.8}[(\phi^2)^{,a}{_{,b}} - (\phi^2)^{,c}{_{,c}}\delta^a{_b}]_{,a} = [\phi^{,a}\phi_{,b} - \phi^{,c}\phi_{,c} \delta^a{_b} + \phi \phi^{,a}{_{,b}} - \phi \phi^{,c}{_{,c}} \delta^a{_b}]_{,a} = 0.\ee
In view of \eqref{2.6} and \eqref{2.8}, we obtain

\be \label{2.9}T_2{^{a}{_{b,a}}} = 0,~~\mathrm{where},~~T_2{^a{_b}} = \phi^{,a}\phi_{,b} - \phi^{,c}\phi_{,c} \delta^a{_b} + \phi T_1{^a{_b}}.\ee
Thus, $T_2{^a{_b}}$ is again an off-shell geometric conserved tensor. Likewise, it is possible to find many other conserved tensors, starting from the scalar function $\phi^n$, where, $n$ is arbitrary, or even for a more general scalar function $f(\phi)$. For example, the integrability condition of the vector $f(\phi)^{,a}$

\be \label{f1} \left[f(\phi)^{,a}\right]_{;b;a} - \left[f(\phi)^{,a}\right]_{;a;b} = 0,\ee
may be expressed as

\be \label{f2} \left[f(\phi)^{,a}\right]_{;b;a} - \left[f(\phi)^{,c}\right]_{;c;a}{\delta^a}_b = 0,\ee
under the change of dummy indices, which finally leads to

\be \label{f3} \left[{f(\phi)^{,a}}_{;b} - {f(\phi)^{,c}}_{;c}{\delta^a}_b\right]_{;a} = 0 = T_f{^{a}{_{b,a}}},\ee
in view of the fact that ${\delta^a}_{b;a} = 0$, and hence

\be \label{f4}{T_f{^{a}}_b} = {f(\phi)^{,a}}_{;b} - {f(\phi)^{,c}}_{;c}{\delta^a}_b,\ee
is a geometric conserved tensor for a general function $f(\phi)$. From expression \eqref{f4}, it is possible to retrieve $T_1{^a{_b}}$, $T_2{^a{_b}}$, and also indefinitely large number of conserved tensors associated with arbitrary function of $f(\phi)$.\\

Nonetheless, the tensor ${{T_2}^a}_b$ is of particular interest, since it is connected with attempts to have a renormalizable theory of $\phi^4$ type of potential (see \cite{JCC, SDJ}, and references therein). The addition of this ${{T_2}^a}_b$ to the usual energy momentum tensor of a scalar field, can also be obtained through Noether's theorem applied to a Lagrangian that differs from the original one by some divergence term, in particular by adding something proportional to $(\phi\phi^c)_{;c}$.

\subsection{Dynamical (on-shell) conserved tensors:}

To find conserved current on-shell, it is required to use the dynamical viz, the Klein-Gordon equation \eqref{2.2}. Let us first multiply equation \eqref{2.4} by $\phi$ and rearrange to obtain,

\be \label{2.10}\phi\phi^{,a}{_{,a,b}} - \phi\phi^{,a}{_{,b,a}} = (\phi\phi^{,a}{_{,a}})_{,b} - (\phi\phi^{,a}{_{,b}})_{,a}  + \phi^{,a}\phi_{,b,a} - \phi_{,b} {\phi^{,a}}_{,a} = 0,\ee
where, we use the symmetry in the index, that is $\phi{_{,a,b}} = \phi{_{,b,a}}$, being integrable. Equation \eqref{2.10} may be expressed as:

\be \label{2.11} (\phi\phi^{,a}{_{,a}})_{,b} = (\phi\phi^{,a}{_{,b}})_{,a} - \phi^{,a}\phi_{,b,a} + \phi_{,b}\phi^{,a}{_{,a}} = (\phi\phi^{,a}{_{,b}})_{,a} - (\phi^{,a}\phi_{,b})_{,a} + 2\phi^{,a}{_{,a}} \phi_{,b}.\ee
Thus finally, we arrive at,

\be \label{2.12} (\phi\phi^{,a}{_{,a}})_{,b} - 2\phi^{,a}{_{,a}} \phi_{,b} = \left[\phi\phi^{,a}{_{,b}}- \phi^{,a}\phi_{,b}\right]_{,a}.\ee
Now since the potential, $V = V(\phi)$, so we may have,

\be \label{2.13} \frac{\partial V}{\partial x^b} = \frac{\partial V}{\partial x^a} \frac{\partial x^a}{\partial x^b} = \frac{\partial}{\partial x^a}\left( V \frac{\partial x^a}{\partial x^b}\right) = (V \delta^a{_{b}})_{,a} = \frac{\partial V}{\partial \phi}\delta^a{_{b}} \phi_{,a}= \frac{\partial V}{\partial \phi} \phi_{,b} = - \phi^{,a}{_{,a}}\phi_{,b},\ee
where, we have used the Klein-Gordon equation \eqref{2.2} in the last step. Therefore, in view of the last two equations \eqref{2.12} and \eqref{2.13} we obtain,

\be \label{2.14} \left[{1\over 2}\phi^{,a}\phi_{,b} - {1\over 2} \phi\phi^{,a}{_{,b}}\right]_{,a} + (V \delta^a{_{b}})_{,a} + \left[{1\over 2} \phi\phi^{,c}{_{,c}}\delta^a{_{b}}\right]_{,a} = T_3{^a}{_b}_{,a}  = 0.\ee
Finally, using the conserved tensor appearing in \eqref{2.6}, we arrive at,

\be \label{2.15} T_3{^a}{_b} = {1\over 2}\phi^{,a}\phi_{,b} + V \delta^a{_{b}} -  {1\over 2}\phi T_1{^a}{_b},\ee
Clearly, $T_3{^a}{_b}$ is an on-shell second rank symmetric conserved tensor.

\subsection{Physical interpretation of the conserved tensors:}

Eliminating $T_1{^a}{_b}$ between \eqref{2.9} and \eqref{2.15} one obtains

\be \label{2.16} T{^a}{_b} = {1\over 2} T_2{^a}{_b} + T_3{^a}{_b} = \phi^{,a}\phi_{,b} - \left({1\over 2} \phi^{,c}\phi_{,c} - V(\phi)\right)\delta^a{_{b}},\ee
which is the stress-energy tensor of the scalar field. Clearly, the additional conserved tensors explored so far, becomes physically meaningful, since, the combination of the on-shell and off-shell conserved tensors leads to the energy-momentum tensor of the scalar field.\\

\noindent
Now, to get some insight of the underlying physics inherent in the off-shell conserved tensor, we take the example of the conserved tensor $T_{1ab}$ for the real scalar field presented in \eqref{2.6}, and show how the non-dynamical (geometric) conserved tensor may be expressed  in terms of the shear, dilation etc.\\

\noindent
In the foliation of the $4$-dimensional space-time into the set of $3$-space section, one can treat time like lines as orthogonal to the family of three-space section (i.e., the tangent to the time like line is orthogonal to the three spatial section).  The scalar field $\phi$ at a given instant of time is distributed in the three-space section and the evolution of the field may be in a direction orthogonal to the three-space section. One may consider the direction of evolution as the direction of gradient of the scalar function and hence may introduce a unit time-like vector,

\be \label{2.17} v_a = \frac{\phi_{,a}}{\sqrt {\phi_{,c}\phi^{,c}}},~~~~\mathrm{where}~~~ v_a v^a = 1.\ee
The direction of evolution will be along the direction of $v_a$ as long as it is time-like. However, one can not predict when the scalar field becomes static and $v_a$ ceases to be a time-like vector and becomes space-like. Nevertheless, in the following analysis we shall consider fields corresponding to which $v_a$ is always a time-like vector. The vector $v_a$ is rotation-free as it is defined from the gradient of a scalar function. Let us define the traceless, symmetric tensor $\sigma_{ab}$  with respect to the time-like vector $v_a$ as:

\be \label{2.18}\sigma_{ab} = v_{a,b} - {\theta\over 3}(\eta_{ab} - v_av_b) - \dot v_a v_b,~~~\mathrm{where},~~~\dot v_a = v_{a,b} v^b;~~~~~~\theta = v^a{,{_a}},~~ \sigma_{ab}v^b=0 .\ee
In the above, $\theta$ determines the rate of dilation of the three-space element locally orthogonal to the vector $v^a$, and $\dot v_a$ is the acceleration vector. Now in view of \eqref{2.17}, we can calculate the acceleration, dilation and shear, which under simplification, lead to,

\bes  \label{2.19}\dot v_a= \frac{\phi_{,a,c}\phi^{,c}}{\phi_{,d}\phi^{,d}} - v_a\frac{\phi_{,c,d}\phi^{,c}\phi^{,d}}{(\phi_{,b}\phi^{,b})^{3\over 2}}; \hspace{0.5 in}\theta = \frac{\phi^{,a}{_{,a}}}{\sqrt{\phi_{,d}\phi^{,d}}}-\frac{\phi_{,c,d}\phi^{,c}\phi^{,d}}{(\phi_{,b}\phi^{,b})^{3\over 2}};\\&\sigma_{ab} = \frac{\phi_{,a,b}}{\sqrt{\phi_{,d}\phi^{,d}}} -(\dot v_av_b + v_a\dot v_b) -v_a v_b\frac{\phi_{,c,d}\phi^{,c}\phi^{,d}}{(\phi_{,b}\phi^{,b})^{3\over 2}} -{\theta\over 3}(\eta_{ab} - v_a v_b).\end{split}\ee
The geometric conserved tensor $T_{1ab}$ obtained in \eqref{2.6} may also be expressed in terms of these kinematical quantities in the following manner:

\be \label{2.20} T_{1ab} = \sqrt{\phi_{,c}\phi^{,c}}[\sigma_{ab} + \dot v_a v_b + v_a\dot v_b] + {\theta\over 3}\sqrt{\phi_{,c}\phi^{,c}}(\eta_{ab} - 4v_a v_b) - \phi^{,c}{_{,c}}(\eta_{ab} - v_a v_b),\ee
which, in view of \eqref{2.18} trivially reduces to,

\be \label{2.21}  T_{1ab} v^b = \sqrt{\phi_{,c}\phi^{,c}}~\dot v_a - \theta \sqrt{\phi_{,c}\phi^{,c}}~v_a.\ee
Further, in view of the expression for the energy density,

\be \label{2.22} E = {1\over 2}\phi_{,a}\phi^{,a} + V(\phi), \ee
and using the field equation \eqref{2.2}, together with equations \eqref{2.19} and \eqref{2.20}, it is possible to express $T_{1ab}$ as,

\be \label{2.23} T_{1ab} v^b = \frac{E_{,a}}{\sqrt{{\phi_{,c}}\phi^{,c}}}.\ee
In view of \eqref{2.21} and \eqref{2.23} we arrive at

\be \label{2.24} \dot{v_{a}}=\frac{E_{,a}}{\phi_{,c}\phi^{,c}}+ \theta v_{a}= \frac{E_{,a}}{2(E-V)} + \theta v_{a},\ee
which effectively gives a dynamical equation of motion, as the acceleration vector is a combination of the gradient of energy and a term proportional to the velocity vector. This appears identical to the Euler hydrodynamical equation in the presence of an equation of state between the pressure and the density in $4$-dimensional space-time. The $4$-force density is $E\dot{v_a}$ in view of \eqref{2.24}. Further under a contraction with $v^a$ the equation \eqref{2.24} reduces to

\be \label{2.25} E_{,a}v^{a}= -\theta (\phi_{,c}\phi^{,c})=-2\theta (E-V).\ee
The relation \eqref{2.25} is similar to the equation of continuity of a fluid (i.e. $E_{,a}v^{a}=- J^a_{~,a}$) assuming energy density to be $E$. Naturally it contains a current density, which is $J^a=2(E-V)v^a$ in \eqref{2.25} as long as $\phi_{,a}\phi^{,a}$ is constant along the 4-velocity vector. Further for any arbitrary time-like vector the expansion scalar $\theta={v^a}_{,a}$ indicates  diverging time-like congruence of geodesics and we have no problem in obtaining the Raychaudhuri equation.  Thus $\theta$ is expansion scalar as long as $v_a$ is a time-like vector or the velocity vector. Again $\theta$ is non-vanishing in \eqref{2.25} if and only if the energy density is changing in the background of space-time provided, $\phi_{,a} \phi^{,a} \ne 0$. Further evolution of $\theta$, i.e. $\theta_{,a}v^{a}$ can be obtained from \eqref{2.4} under contraction with $\phi^{,b}$ and using \eqref{2.19} as:

\be \label{2.26} \theta_{,a} v^a + {\theta^2\over 3} + \sigma_{ab} \sigma^{ab} - \dot v^a{_{,a}} = 0.\ee
Above equation \eqref{2.26} determines the evolution of the rate of dilation of the scalar field in the background of flat space-time. The divergence of the acceleration vector $\dot v^a{_{,a}}$ may be positive, negative or zero, depending upon the state of the scalar field. The magnitude of the shear tensor $\sigma_{ab}$ is $|\sigma|$, where, $\sigma^2 = {1\over 2} \sigma_{ab}\sigma^{ab} \ge 0$. It is important to mention that equation \eqref{2.26} is similar to the Raychaudhuri equation \cite{AKR} of relativistic cosmology apart from the source term that appears in General Theory of Relativity.

\subsection{Summary:}

The integrability condition of a vector obtained from the gradient of a scalar function appears as the fundamental condition to obtain conserved tensors in flat space-time. A suitable linear combination of the off-shell (geometric) and the on-shell (dynamical) conserved tensors corresponding to the real scalar field, lead to the stress-energy tensor. The rate of dilation ($\theta$) of the scalar field is non-vanishing, provided the energy-density is changing, and $\phi_{,a}\phi^{,a} \ne 0$.

\section{Complex scalar field in flat (Minkowski) space-time:}

The Lagrangian density for a complex scalar field in flat space-time is expressed in the form

\be \label{3.1}\mathcal{L} =  \phi^*_{,\alpha}\phi^{\alpha} - V(\phi^*,\phi).\ee
The field equations are,

\be \label{3.2}\phi^{*,a}{_{,a}} + \frac{\partial V}{\partial \phi} = 0; ~~\mathrm{and},~~\phi^{,a}{_{,a}} + \frac{\partial V}{\partial \phi^*} = 0.\ee
The conserved (${T^{ab}}_{,b} = 0$) stress-energy tensor is:

\bes\label{3.3} T_{ab} = {\partial \mathcal{L}\over \partial(\partial^a \phi)}\partial_b \phi + {\partial \mathcal{L}\over \partial(\partial^a \phi^*)}\partial_b \phi^* - \eta_{ab}\mathcal{L}\\&
\hspace{4.4 mm}= \phi_{,a}\phi^*_{,b} + \phi^*_{,a}\phi_{,b} -{1\over 2}\eta_{ab}\eta^{cd}(\phi_{,c}\phi^*_{,d} + \phi^*_{,c}\phi_{,d}) + \eta_{ab}V(\phi,\phi^*).\end{split}\ee
Further, the conserved current reads as,

\be\label{3.4} J_a = \phi^* \partial _a\phi - (\partial_a\phi^*)\phi.\ee

\subsection{Non-dynamical (off-shell) conserved tensors:}

Using the contracted form of the Ricci identity \eqref{2.3} for the vectors $\phi_{,a}$ and $\phi^*{_{,a}}$ we have

\be\label{3.5} \phi^{,a}{_{,a,b}} = \phi^{,a}{_{,b,a}};~~~\mathrm{and},~~~\phi^{*,a}{_{,a,b}} = \phi^{*,a}{_{,b,a}},\ee
which imply that for a scalar field, the order of the covariant derivatives is immaterial. The conserved tensors are therefore,

\be \label{3.6}T_{4ab} = \phi_{,a,b} - \phi^{,c}{_{,c}}\eta_{ab};~~~~~\mathrm{and},~~~~~T_{5ab} = \phi^*{_{,a,b}} - \phi^{*,c}{_{,c}}\eta_{ab}.\ee
Since $\phi$ and $\phi^*$ are integrable, ie. $\phi_{,a,b} = \phi_{,b,a}$ and $\phi^*{_{,a,b}}=\phi^*{_{,b,a}}$, so the conserved tensors $T_{4ab}$ and $T_{5ab}$ are symmetric. Further, since $\phi^*\phi$ is a scalar, so we also have

\be \label{3.7}(\phi^*\phi)^{,a}{_{,a,b}} - (\phi^*\phi)^{,a}{_{,b,a}} = 0,\ee
in view of which one can construct yet another conserved tensor, viz.

\be\label{3.8} T_{6ab} = (\phi^*\phi)_{,a,b} - (\phi^*\phi)^{,c}{_{,c}} \eta_{ab}.\ee
The above form of the conserved current may be simplified by the use of \eqref{3.6} to take the following form,

\be \label{3.9} T_{6ab} = \phi^*{_{,a}}\phi_{,b} + \phi^*{_{,b}}\phi_{,a} - 2\phi^*{_{,c}}\phi^{,c}\eta_{ab} +
\phi^* T{_{4ab}} + \phi T{_{5ab}},\ee
which is again a second rank symmetric off-shell (geometric) conserved tensor for an arbitrary complex scalar field.

\subsection{Dynamical (on-shell) conserved tensors:}

Multiplying the first of equations \eqref{3.5} by $\phi^*$ and rearranging the terms and the tensor indices, we obtain

\be\label{3.10} \Big[\phi^{*,a}\phi_{,b} - \phi^* \phi^{,a}{_{,b}} + \phi^* \phi^{,c}{_{,c}} \delta^a{_b}\Big]_{,a} - \phi^{,a}{_{,a}} \phi^*{_{,b}} - \phi^{*,a}{_{,a}} \phi{_{,b}}= 0.\ee
Since the potential $ V = V(\phi, \phi^*)$, so

\be\label{3.11} {\partial V\over \partial x^b} = {\partial V\over \partial \phi}{\partial \phi\over \partial x^b} + {\partial V\over \partial \phi^*}{\partial \phi^*\over \partial x^b} = - \phi^{*,a}{_{,a}} \phi{_{,b}} -  \phi^{,a}{_{,a}} \phi^*{_{,b}},\ee
in view of field equations \eqref{3.2}. Therefore, in view of equations \eqref{3.10} and \eqref{3.11}, we obtain

\be\label{3.12} T_7{^a}{_{b,a}} = 0, ~~~\mathrm{where},~~~ T_7{^a}{_b} = \phi^{*,a}\phi_{,b} + V \delta^a{_b} - \phi^* T_4{^a}{_b}.\ee
Hence, $T_7{^a}{_b}$ is an on-shell (dynamical) conserved tensor and is different from the energy-momentum tensor of the complex scalar field. Likewise, multiplying the second of equations \eqref{3.5} by $\phi$ and rearranging the terms and the tensor indices, one can also obtain another conserved tensor in the form,

\be \label{3.13} T_8{^a}{_{b,a}} = 0, ~~~\mathrm{where},~~~ T_8{^a}{_b} = \phi^{,a}\phi^*{_{,b}} + V \delta^a{_b} - \phi T_5{^a}{_b}.\ee
Finally, in view of equations \eqref{3.9}, \eqref{3.12} and \eqref{3.13}, we can construct yet another conserved current in the form,

\be\label{3.14} T_9{^a}{_{b,a}} = 0, ~~\mathrm{where},~~ T_{9ab} = {1\over 2}\big(T_{6ab}+ T_{7ab}+ T_{8ab}\big) = \phi^*{_{,a}}\phi_{,b} + \phi_{,a}\phi^*{_{,b}} -(\phi^*{_{,c}} \phi^{,c} - V) \eta{_{ab}}.\ee
The conserved tensor $T_{9ab}$ may be cast in the form $T_{ab}$ \eqref{3.3}, and thus the energy-momentum tensor of the complex scalar field has also been retrieved from the combination of non-dynamical (off-shell, viz. $T_{6ab}$) and dynamical (on-shell, viz. $T_{7ab}~ \& ~T_{8ab}$) conserved tensors.

\subsection{More information from complex scalar fields:}

Now, to explore additional information from the complex scalar field, we substitute $\phi = \alpha e^{iS}$ in the Lagrangian \eqref{2.1}, where, both the amplitude $\alpha = \alpha(x^a)$ and the phase $S = S(x^a)$ are functions of the space-time coordinates. Thus, the field Lagrangian \eqref{2.1} is transformed to,

\be \label{3.15} \mathcal{L} = \eta^{ab}(\alpha_{,a}\alpha_{,b} + \alpha^2 S_{,a} S_{,b}) - m^2 \alpha^2,\ee
where, we have chosen the scalar potential $V(\phi, \phi^*)  = m^2 \phi^*\phi$. The field equation \eqref{3.2} then takes the form:

\be \label{3.16} \eta^{ab} (\alpha_{,a,b} - \alpha S_{,a}S_{,b}) + m^2 \alpha = 0,\ee
while the continuity equation, viz.,

\be \label{3.17} (\eta^{ab} \alpha^2 S_{,a})_{,b} = 0,\ee
essentially gives the conserved current \eqref{3.4}. The stress-energy tensor \eqref{3.3} in view of $\phi = \alpha e^{iS}$ reads as,

\be \label{3.18} T_{ab} = 2 (\alpha_{,a}\alpha_{,b} + \alpha^2 S_{,a} S_{,b}) + \eta_{ab}m^2 \alpha^2 - \eta_{ab}\eta^{cd}(\alpha_{,c}\alpha_{,d} + \alpha^2 S_{,c} S_{,d}).\ee
Now to incorporate the conserved current \eqref{3.17} into the field equation \eqref{3.16}, we divide equation \eqref{3.16} throughout by $\alpha$ and then differentiate with respect to $x^c$. Further, using symmetry property of the indices along with equation \eqref{3.17}, we finally obtain,

\be\label{3.19} L_{cb}{^{,b}} = \eta^{ab}(\alpha_{,c}\alpha_{,b} - \alpha\alpha_{,b,c}  + 2\alpha^2 S_{,b} S_{,c})_{,a} = 0.\ee
This is quite different from the above form of the stress-energy tensor \eqref{3.18}, and therefore is a new conserved second rank tensor. In fact, for constant amplitude $\alpha$, i.e. for plane wave solution of $\phi$, if we eliminate the mass term between the field equation \eqref{3.16} and the stress-energy tensor \eqref{3.18}, then one obtains,

\be \label{3.20} T_{ab} = 2 \alpha^2 S_{,a} S_{,b} = L_{ab}.\ee
Thus, for constant amplitude, the stress energy tensor is identical to the new conserved tensor $L_{ab}$. However, as mentioned, $L_{ab}$ ceases to be the stress-energy tensor, for variable amplitude $\alpha$, i.e. under the departure from plane-wave solution. To understand the difference better, let us express equation \eqref{3.19} in terms of $\phi$ and $\phi^*$ as,

\be \label{3.21} L_{cb}{^{,b}} = \frac{1}{2}\eta^{ab}(\phi_{,b}\phi^*{_{,c}} + \phi_{,c}\phi^*{_{,b}} - \phi\phi^*{_{,b,c}} - \phi^*\phi{_{,c,b}})_{,a}.\ee
Clearly, $L_{cb}$ contains upto second order derivatives of the field variable, while stress energy tensor \eqref{3.3} contains up to first order. Of course in principle, it is possible to express $T_{ab}$ in terms of the second order derivatives of the field variable, by eliminating the mass term between $T_{ab}$ and the field equation \eqref{3.16}. Even then however, it is not the same as $L_{ab}$. This is simply because, on the contrary, it is not possible to express $L_{ab}$ in terms of first derivative terms only. Thus, treating amplitude as a function of space-time coordinates, an additional symmetry of the Lorentz invariant Lagrangian, associated with the new conserved tensor $L_{ab}$ emerges. The importance of such a conserved tensor is required to understand better. The reason being: the stress-energy tensor is supposed to be the unique second rank conserved quantity apart from a divergence-free term.  Conservation of $L_{ab}$ contradicts this very uniqueness of the stress-energy tensor. Therefore, just like the stress-energy tensor and the conserved charge, one would like to incorporate $L_{ab}$ as a source term of certain field equation. In fact, the field to which $L_{ab}$ should be coupled may be known only after determining the symmetry of the Lagrangian associated with $L_{ab}$. Nonetheless, in the following section we show that even without having a precise knowledge of the associated symmetry, it is in principle possible to find the field equation for which $L_{ab}$ may act as a source.

\section{The magic with a complex singlet in curved space-time:}

Conserved quantities are usually treated as the sources of some field theory. For example, one would like to replace $\eta_{ab}$ by $g_{ab}$ in the expression for the stress-energy tensor \eqref{3.3} in curved space-time and then use it as the source of the gravitational field. The conserved current \eqref{3.4} on the contrary, may be treated as the source of Maxwell's equation with due care \cite{Max}. However, as we see below, the situation is much more involved in curved space-time, rather than trivially generalizing equations \eqref{3.19} or \eqref{3.21} from flat space to the curved space, simply by replacing the ordinary derivatives (commas) with the covariant derivatives (semicolons) along with the replacement of $\eta_{ab}$ by $g_{ab}$. Further, for any scalar function $\phi(x^a)$ we have $\phi_{;a}=\phi_{,a}$. \\

To get the dynamics of a scalar field under the influence of gravity, we have to consider contribution of curved space-time in the dynamics of the scalar field. We can achieve this simply by introducing the Ricci identity satisfied by a vector field obtained from the gradient of the scalar field $\phi$. In the process, the effect of gravitational field automatically appears just by invoking general covariance only, and much more than the generalized version of the energy momentum stress tensor of the scalar field \eqref{3.20} in curved space-time (for constant amplitude), is found. Let us start with the general covariant Klein-Gordon equations for a complex scalar field in curved space-time, viz.,

\be \label{4.1} (g^{ab}\phi_{,b})_{;a} + m^2 \phi = 0, ~~~~~\mathrm{and}~~~~~(g^{ab}\phi^*{_{,b}})_{;a} + m^2 \phi^* = 0.\ee
Taking derivative of \eqref{4.1} with respect to $x^{c}$ and using Ricci identity $(R_{ca}\phi^{,a} = {\phi^{,a}}_{;c;a}-{\phi^{,a}}_{;a;c} )$, we obtain,

\be \label{4.2}-\phi^* R_{ca}\phi^{,a} + \phi^*{\phi^{,a}}_{;c;a} + m^2 \phi^*\phi_{,c} = 0,\ee
where, we have multiplied throughout by $\phi^*$. Arranging the above equation, we may write:

\be \label{4.3}-(\phi^* \phi R_{ca})^{;a} + \phi^* \phi{ R_{ca}}^{;a} + \phi^{*,a} \phi R_{ca} + (\phi^*{\phi^{,a}}_{;c})_{;a} - {\phi^{*}}_{,a}{\phi^{,a}}_{;c}+ m^2 \phi^*\phi_{,c} = 0.\ee
Now using `contracted Bianchi identity' $({R_{ca}}^{;a} = {1\over 2} R_{,c})$ and rearranging, we obtain:

\be \label{4.4}-(\phi^* \phi R_{ca})^{;a} + {1\over 2}(\phi^* \phi R {\delta^{a}}_{c})_{;a} - {1\over 2}(\phi^* \phi)_{,a}R {\delta^{a}}_{c} + \phi \phi^{*,a} R_{ca} + (\phi^*{\phi^{,a}}_{;c})_{;a} - {\phi^{*}}^{,a}{\phi^{,c}}_{;a}+ m^2 \phi^*\phi_{,c} = 0,\ee
since for a scalar, derivatives commute. Using Ricci identity in the fourth term, we can now arrange the above equation as:

\be \label{4.5}\begin{split}\left[\phi^* \phi \Big({R^{a}}_c - {1\over 2}R {\delta^{a}}_{c}\Big)\right]_{;a} &+ {1\over 2}(\phi^* \phi)_{,a}R {\delta^{a}}_{c} \\&= -\phi{\phi^{*,a}}_{;a;c} + \phi{\phi^{*,a}}_{;c;a} + (\phi^*{\phi^{,a}}_{;c})_{;a} - ({\phi^{*,a}}{\phi_{,c}})_{;a} + {\phi^{*,a}}_{;a}\phi_{,c} + m^2 \phi^*\phi_{,c}.\end{split}\ee
Since, the last two terms vanish due to the field equation, we can further arrange the above equation as:

\be \label{4.6}\begin{split}\left[\phi^* \phi \Big({R^{a}}_c - {1\over 2}R {\delta^{a}}_{c}\Big)\right]_{;a} &+ {1\over 2}(\phi^* \phi)_{,a}R {\delta^{a}}_{c} \\&=  m^2\phi\phi^*_{,c} + (\phi{\phi^{*,a}}_{;c})_{;a} - \phi_{,a}{\phi^{*,a}}_{;c} + (\phi^*{\phi^{,a}}_{;c} + {\phi^{*,a}}{\phi_{,c}})_{;a}.\end{split}\ee
Again, since derivatives commute for scalar, we can rearrange the third term on the right hand side to obtain:

\be \label{4.7}\begin{split}\left[\phi^* \phi \Big({R^{a}}_c - {1\over 2}R {\delta^{a}}_{c}\Big)\right]_{;a} &+ {1\over 2}(\phi^* \phi)_{,a}R {\delta^{a}}_{c} \\& = m^2\phi\phi^*_{,c} + (\phi{\phi^{*,a}}_{;c})_{;a} - (\phi^{,a}{\phi_{*,c}})_{;a} + {\phi^{,a}}_{,a}\phi^*_{,c} + (\phi^*{\phi^{,a}}_{;c} + {\phi^{*,a}}{\phi_{,c}})_{;a}.\end{split}\ee
First term on the right hand side gets cancelled with the fourth, and finally the above expression \eqref{4.7} simplifies to:

\be \label{4.8}\left[\phi^* \phi \Big({R^{a}}_c - {1\over 2}R {\delta^{a}}_{c}\Big)\right]_{;a} + {1\over 2}(\phi^* \phi)_{,a}R {\delta^{a}}_{c} = \left[ \phi^*{\phi^{,a}}_{;c} + \phi{\phi^{*,a}}_{;c} -{\phi^{*,a}}{\phi_{,c}} - \phi^{,a}{{\phi^*}_{,c}}\right]_{;a} .\ee
Unlike the flat space situation, we do not obtain a conserved quantity here. Nevertheless, some interesting features of the above equation \eqref{4.8} may be discussed. Firstly, although coupling between the scalar field and the curvature has not been considered from the beginning, the curvature terms appear automatically on the left hand side of the above equation, through Ricci and Bianchi identities. This implies that general covariance is sufficient to introduce curvature. Next, upon substituting $\phi = \alpha e^{iS}$ and $\phi^* = \alpha e^{-iS}$, \eqref{4.8} reduces to,

\be\label{4.9}\Big[\alpha^2 \Big(R_{c}{^a} - {1\over 2} \delta_c{^a} R\Big)\Big]_{;a} + \alpha\alpha_{,c} R = -2(\alpha_{,c}\alpha^{,a} - \alpha\alpha^{,a}{_{;c}} + 2 \alpha^2 S^{,a}S_{,c})_{;a},\ee
and one can note that the right hand side is simply a trivial generalization of equation \eqref{3.20}, obtained in flat space-time, in the sense that ordinary derivative has been replaced by covariant derivative. Thus, in the Ricci flat space, we obtain $L_c{^a}{_{;a}} = 0$, which is a trivial generalization of equation \eqref{3.20}, as mentioned, while additionally we obtain the left hand side. Finally, one can observe that for constant amplitude the coupling in equation \eqref{4.9} disappears, and one can recover an equation similar to Einstein's equation, viz.,

\be\label{4.10} R_c{^a} - {1\over 2} \delta_c{^a} R = -\kappa S^{,a}S_{,c},\ee
apart from a divergence term, whose right hand side has already been identified as the stress-energy tensor, in equation \eqref{3.20} of the previous section. Further the strength of the gravitational coupling (i.e. effective gravitational constant) between the geometry and the matter in \eqref{4.10} is fixed for constant $\alpha$. It is important to mention that, in the Einstein field equations, the contracted Bianchi identity ensures consistency with the vanishing divergence of the matter stress-energy tensor. Same result has been administered here too, but only for constant amplitude (plane wave solution). Over and above, starting from general covariance and constant amplitude, it is possible to retrieve Einstein's equation. It should be mentioned that starting with constant amplitude from the very beginning, one ends up with

\be \label{4.11}  (S^{,a}S_{,c})_{;a} = 0 = T^a{_c}{_{;a}},\ee
so that equivalence principle may then be used to write Einstein's equation as

\be\label{4.12} R_c{^a} - {1\over 2} \delta_c{^a} R = -\kappa S^{,a}S_{,c},\ee
since $(R_c{^a} - {1\over 2} \delta_c{^a} R)_{;a} = 0$ and $\kappa$ is a constant. In a nut-shell, for $\alpha = \alpha(x^a)$, a non-minimally coupled equation \eqref{4.8} as well as \eqref{4.9} emerges, which is not divergence free. For constant amplitude $\phi^*\phi = \alpha^2 = \mathrm{constant}$, however, such problem disappears, and  minimally coupled field equation emerges.

\section{Real scalar field in curved spacetime and similarity of conserved tensor with generalized Brans-Dicke field equation:}

Conserved quantities in curved spacetime may be found for an arbitrary scalar function $f(\phi)$, assuming a slow variation of the real scalar field $\phi(x^a)$. To explore this fact, we start from the Ricci identity of a vector built from an arbitrary function of $\phi$, viz. $f(\phi)^{;a}$, which is:

\be\label{4.16}
{f^{;a}}_{;c;a}-{f^{;a}}_{;a;c} = R_{ca}f^{;a},
\ee
and rearrange different terms using contracted form of Bianchi identity ${R_{ac,}}^{;a}= \frac{1}{2}R_{;a} \delta ^a_c$  as
\be\label{4.17}
{f^{;a}}_{;c;a}-(\Box{f})_{;a}\delta^a_c =(R_{ca}f)^{;a}-R_{ca}^{~~;a}f= (R_{ca}f)^{;a}-\frac{1}{2} R_{;a}\delta^a_c f= (R_{ca}f)^{;a}-\Big(\frac{1}{2}R g_{ac} f\Big)^{;a}+ \frac{1}{2}R g_{ac} f^{;a}.
\ee
Equation \eqref{4.17} may now be conveniently written as,
\be\label{4.18}
\left[f\Big(R_{ca}-\frac{1}{2}R g_{ca}\Big)\right]^{;a} +\frac{1}{2}R g_{ca} f^{;a} = \Big[f_{;c;a}-\square f g_{ca}  \Big]^{;a},
\ee
The second term in the left side of \eqref{4.18} may be expressed as $\frac{1}{2}Rf_{;c}=\frac{1}{2}R \frac{\partial f}{\partial \phi} \phi_{;c}$. Now assuming $f(\phi)$ to be such a slowly varying function, so that the contribution of this term is neglected in comparison with the rest, we have

\be\label{4.19}
\left[f(\phi)\Big(R_{ca}-\frac{1}{2}R g_{ca}\Big) \right]^{;a} \approx  \Big[\{f(\phi)\}_{;c;a}-\square \{f(\phi)\} g_{ca}  \Big]^{;a}.
\ee
Thus, we obtain

\be\label{4.19}
f(\phi)\Big[R_{ca}-\frac{1}{2}R g_{ca}\Big]  =  \{f(\phi)\}_{;c;a}-\square \{f(\phi)\} g_{ca} -\Lambda g_{ca} + \kappa T_{ca},
\ee
where, $\Lambda$ and $\kappa$ are constants, and the energy-momentum tensor of an arbitrary field (matter) $T_{ca}$ satisfies the condition ${T_{ca}}^{;a}=0$. Equation \eqref{4.19} is interesting as it determines the curvature of a space-time, that emerges from an arbitrary scalar field $f(\phi)$, a conserved tensor $T_{ac}$ and a constant $\Lambda$ term. Now assuming the function as $f(\phi)=\phi^n$ the equation \eqref{4.19} reduces to

\be\label{4.20}
\phi\Big[R_{ca}-\frac{1}{2}R g_{ca}\Big]= n \phi_{;c;a}-n\square\phi g_{ca}+ \frac{n(n-1)}{\phi}\phi_{;c}\phi_{;a}-n(n-1)\frac{\phi_{;d}\phi^{;d}}{\phi}g_{ca}- \Lambda \phi^{1-n}g_{ca} + \kappa\phi^{1-n}T_{ca},
\ee
where $n$ is a constant, and the variable ${\phi^{-1}}$ acts as an effective gravitational constant. It is important to mention that the equation \eqref{4.20} is a reminiscent of Bianchi identity and the Ricci identity of covariant theory. Further, in deriving \eqref{4.20}, other than general covariance, nowhere in the computation we use the principle of equivalence or the Klein-Gordon equation. Thus, simply invoking the Ricci identity and the Bianchi identity a scalar field appears to induce a curvature in the space-time in its general covariant form through equation \eqref{4.20}. Note that, in deriving \eqref{4.20} there is no need to introduce the curvature tensor through the Lie bracket of any vector fields. The Mach principle \cite{mach} incorporated in this manner is an alternative idea of the principle of equivalence in the covariant formulation of the theory of gravity.\\

\noindent
Now, the Brans-Dicke-Jordan field equation for the scalar field $\chi$ reads as

\be\label{4.21}
\chi\Big[R_{ca}-\frac{1}{2}R g_{ca}\Big]=  \chi_{;c;a}-\square\chi g_{ca}+ \frac{\omega(\chi)}{\chi}\chi_{;c}\chi_{;a}-\omega(\chi)\frac{\chi_{;d}\chi^{;d}}{2\chi}g_{ca}- \frac{1}{2}V( \chi)g_{ca} + \frac{\kappa}{2}
T_{ca}.\ee
Clearly, the form of equation \eqref{4.20} is quite similar to the Brans-Dicke-Jordan field equation \eqref{4.21} of the generalized scalar tensor theory, where gravitational interaction is mediated not only by the metric tensor $g_{ab}$, but also with a scalar field $\phi$. An important point to mention is: the Brans Dicke field $\chi$ in \eqref{4.21} may not in general be a slowly varying function, while we found \eqref{4.20} neglecting the contribution of the term $R \frac{\partial f}{\partial \phi} \phi_{;c}$. Despite certain similarities, differences in the two equations are also apparent, particularly in the context of coupling parameters. Further, the prefactors of $T_{ac}$ are different and the potential $V(\phi)$ in \eqref{4.20} is not arbitrary, as in \eqref{4.21}.\\

\section{A class of geometric conserved tensors starting from Ricci scalar:}

The field equations \eqref{4.19}, or the equations obtained by introducing the Bianchi identity and Ricci identity for $f(\phi)^{;\mu}$ are not unique example in the theory of gravitation, rather these identities also yield field equations for the modified theory of gravity, as we explore in this section. Let us start from the $n^{\mathrm{th}}$ power of Ricci scalar $R$. Using Ricci identity for the vector $(R^n)^{;\mu}$,

\be\label{5.1} {(R^n)^{,\mu}}_{;\nu;\mu} - {(R^n)^{,\mu}}_{;\mu;\nu} = R_{\mu\nu} (R^n)^{,\mu},\ee
and expressing it as total derivative, one obtains,

\be \label{5.2}\begin{split} {(R^n)^{,\mu}}_{;\nu;\mu} - \left[{(R^n)^{,\alpha}}_{;\alpha}{\delta^\mu}_\nu\right]_{;\mu} & = \left(R_{\mu\nu} R^n\right)^{,\mu} - (R_{\mu\nu})^{;\mu} R^n
= \left({R^\mu}_\nu R^n\right)_{,\mu} - {R_{;\nu}\over 2} R^n\\&
= \left({R^\mu}_\nu R^n\right)_{,\mu} - {{(R^{n+1})}_{;\nu}\over 2(n+1)} = \left({R^\mu}_\nu R^n\right)_{,\mu}- \left[{R^{n+1}\over 2(n+1)}{\delta^\mu}_\nu\right]_{;\mu},\end{split}\ee
where we have used contracted Bianchi identity. Thus one arrives at,

\be \label{5.3}  \left[{(R^n)^{,\alpha}}_{;\alpha}{\delta^\mu}_\nu - {(R^n)^{,\mu}}_{;\nu} + {R^\mu}_\nu R^n - {R^{n+1}\over 2(n+1)}{\delta^\mu}_\nu\right]_{;\mu} = 0.\ee
Hence, we obtain a class of geometric conserved tensors ${T(n)^\mu}_{\nu}$ of rank two depending on $n$ as,

\be \label{5.4} {T(n)^\mu}_{\nu} = R^n\left[{R^\mu}_\nu - {R\over 2(n+1)}{\delta^\mu}_\nu\right] + {(R^n)^{,\alpha}}_{;\alpha}{\delta^\mu}_\nu - {(R^n)^{,\mu}}_{;\nu},\ee
where, $\left({T(n)^\mu}_{\nu}\right)_{;\mu} =0$. Note that, equation \eqref{5.4} is essentially an outcome of the Ricci identity for the vector $(R^n)^{;\mu}$ and the Bianchi identity, where, $n$ is arbitrary (including fractional values), except $n \ne -1$. Interestingly, one can now set $n = 0$, to obtain Einstein's tensor as:

\be \label{5.5} T(0)_{\mu\nu} = R_{\mu\nu} - {1\over 2} g_{\mu\nu}R = G_{\mu\nu}.\ee
Further, setting $n = 1$, the tensor associated with $R^2$ term is found \cite{RM}, viz:

\be \label{5.6} {T(1)^\mu}_{\nu} = R\left[{R^\mu}_\nu  - {R\over 4}{\delta^\mu}_\nu\right] - {R^{,\mu}}_{;\nu} + {R^{,\alpha}}_{;\alpha}{\delta^\mu}_\nu. \ee
Likewise for $n = 2$, the tensor in connection with $R^3$ is retrieved as:

\be \label{5.7}  {T(2)^\mu}_{\nu} = R^2\left[{R^\mu}_\nu  - {R\over 6}{\delta^\mu}_\nu\right]  - {(R^2)^{,\mu}}_{;\nu} + {(R^2)^{,\alpha}}_{;\alpha}{\delta^\mu}_\nu,\ee
and so on. At this end, one can introduce equivalence principle to associate an energy momentum tensor on the right hand side of \eqref{5.4} to find respective field equations.\\

However, to find conserved tensor for a combination of different powers of curvature scalars, particularly for the modified theory of gravity, one has to start with a more general form viz., $F(R)$. Using Ricci identity for the vector $F(R)^{;\mu}$,

\be\label{F1} {F(R)^{,\mu}}_{;\nu;\mu} - {F(R)^{,\mu}}_{;\mu;\nu} = R_{\mu\nu} F(R)^{,\mu},\ee
and expressing it as total derivative, we obtain,

\be\label{F2} \left[{F(R)^{,\mu}}_{;\nu} - {F(R)^{,\alpha}}_{;\alpha}{\delta^\mu}_\nu\right]_{;\mu} = \left[{R^\mu}_\nu F(R)\right]_{;\mu} - F(R) {R_{\mu\nu}}^{;\mu} = \left[{R^\mu}_\nu F(R)\right]_{;\mu} - {F(R)\over 2} R_{,\nu},\ee
which may further be rearranged as

\be\label{F3} \left[F(R){R^\mu}_\nu + {F(R)^{,\alpha}}_{;\alpha}{\delta^\mu}_\nu -  {F(R)^{,\mu}}_{;\nu} \right]_{;\mu} = \left[{F(R)\over 2} R\right]_{,\nu} - {1\over 2}F(R)_{,\nu}R.\ee
Thus finally we obtain

\be\label{F4} \left[F(R){R^\mu}_\nu  - {1\over 2}R F(R) {\delta^\mu}_\nu  - {F(R)^{,\mu}}_{;\nu} + {F(R)^{,\alpha}}_{;\alpha}{\delta^\mu}_\nu  \right]_{;\mu} + {1\over 2}F(R)_{,\nu}R = 0.\ee
Clearly, the last term cannot be expressed as a total derivative term in general. Nonetheless, considering

\be \label{FR} F(R) = \sum_{n = 0} f_n R^n,\ee
where $F_n$ is a constant, we can express the last term of the equation \eqref{F4} as,

\be\label{Fnu} {1\over 2}R F(R)_{,\nu} = {1\over 2}\sum_{n = 0}^nnf_n  R^n R_{,\nu} = {1\over 2}\sum_{n = 0}^n \left({n\over n+1} f_n R^{n+1}\right)_{,\nu}.\ee
As a result, equation \eqref{F4} now reads as,

\be\label{F5}\left[\sum_{n = 0} f_n \left\{R^n\Big({R^\mu}_\nu  - {1\over 2}R {\delta^\mu}_\nu\Big)  - {\left(R^n\right)^{,\mu}}_{;\nu} + \left(\Box{R^n}\right){\delta^\mu}_\nu + {1\over 2} \left({n\over n+1} R^{n+1}{\delta^\mu}_\nu\right) \right\}\right]_{;\mu}  = 0.\ee
Thus, we obtain a class of conserved tensors for all possible combinations of different powers of the scalar curvature terms except for $n = -1$. Choosing $n = 0$, $n = 1$ and $n =2$, conserved tensors \eqref{5.5}, \eqref{5.6} and \eqref{5.7} respectively are found. On the contrary, if $n$ runs form $n = 0 ~\mathrm{to}~ 2$, then we obtain conserved tensor for $F(R) = \alpha R + \beta R^2 + \gamma R^3$, where, $\alpha = f_0$, $\beta = f_1$ and $\gamma = f_2$, as:

\be\label{F7}\begin{split} &\Bigg[\alpha\left\{R_{\mu\nu} - {1\over 2} R g_{\mu\nu}\right\} + \beta\left\{R\left(R_{\mu\nu}  - {1\over 4} R g_{\mu\nu}\right) - R_{;\mu;\nu} + \Box R g_{\mu\nu}\right\}\\&
\hspace{1.14 in} + \gamma\left\{R^2\left(R_{\mu\nu} - {1\over 6}R g_{\mu\nu}\right) - (R^2)_{;\mu;\nu} + \Box(R^2)g_{\mu\nu} \right\}\Bigg]_{;\mu} = 0.\end{split}\ee
One can now associate equivalence principle and call upon an appropriate energy-momentum tensor for arbitrary matter field, satisfying ${{T^\mu}_{\nu}}_{;\mu} = 0$, to obtain gravitational field equations as well.

\section{Conclusion}

In this work we use geometric criteria to reproduce dynamical and non-dynamical conservation laws depending on whether the equations of motion are necessary for their satisfaction or not. We start from the Ricci identity and explore the possibility of constructing such tensors in the context of scalar field theory both in flat and in curved manifolds. In the penultimate section we also derive several tensors with the property of being conserved off shell and which can be seen as reproducing the equations of motion of a power law $F(R) \propto R^n$ gravity. Even though it is well known that one can add quantities that are trivially conserved in the energy-momentum tensor of a theory, the derivation of such tensors through purely geometric arguments is certainly useful.\\

To be more precise, we have explored a host of off-shell conserved currents, starting from the integrability condition, for a real as well as complex scalar field in flat space-time. Some on-shell conserved currents have also been explored for the associated of Klein-Gordon equation. An important outcome is the retrieval of the stress-energy tensor of the scalar field \eqref{2.16} and \eqref{3.14}, under linear combination of the on-shell and off-shell conserved currents. Additionally, Euler hydrodynamical equation and the rate of dilation have been found in view of the off-shell conserved currents. The on-shell conserved current for a complex scalar field, has also been identified with the stress-energy tensor, for constant amplitude.\\

Although, trivial generalization of these conservation laws under replacement of ordinary derivative (comma) by covariant derivative (semicolon) reflects zero divergence ${T^{\mu\nu}}_{;\nu} =0$ of the energy-momentum tensor, yet in curved space-time, it does not imply a true conservation law as it does in special relativity. This is because, in curved space-time there is the gravitational energy, that is not included in the energy-momentum tensor. We therefore also attempted to incorporate gravity, using Ricci and Bianchi identities in the field equations.\\

In curved space-time, Ricci identity is the integrability condition for the existence of a metric having a given Riemann curvature tensor. Bianchi identity, on the contrary, is an ingredient of the curvature tensor $R_{abcd}$, or equivalently, a gauge equation in the Riemannian space. In other words, if one asks which tensors on a four-dimensional manifold are Riemann tensors of Lorentz metrics, then there are certain well-known integrability conditions which must be fulfilled, namely the Bianchi and Ricci identities \cite{Ran}. The Ricci identity relates the commutator of two successive covariant derivatives to the Riemann tensor and can be considered to be an implicit definition of the Riemann curvature tensor. The Ricci identity appears naturally during the process for a repeated covariant differentiation, which is not a commutative operation. The basic Ricci identity is simply one of the many ways of defining curvature. It applies universally to connections and curvatures of all bundles. It is possible to find solutions of Einstein’s equations with specific curvature properties by solving the Bianchi and Ricci identities. Such methods have been used earlier by Ellis \cite{Ellis} and Held \cite{Held1, Held2}. \\

We introduce curvature using Ricci and Bianchi identities, and unlike the flat space-time case, ended up with expression \eqref{4.8} and \eqref{4.18}, without any conservation law. However, the fact that the contracted Bianchi identity ensures consistency with vanishing divergence of the stress-energy tensor for matter in the case of Einstein's field equations for general relativity, has been approved here too, only for constant amplitude, in which case Einstein's equation has been retrieved with appropriate gravitational coupling. In particular, we have identified the field equations more-or-less with Brans-Dicke-Jordan scalar tensor theory of gravity, for slow variation of the scalar field, without using equivalence principle in our computation. More precisely, we show the application of our technique in gravity-fluid duality as well. Additionally, starting from the Ricci scalar ($R^n$), conserved tensors associated with different powers of $n$, starting from the Einstein's tensor have been found. Last but not the least, conserved tensor for a modified $[F(R)]$ theory of gravity has been explored, following this simple technique. One can associate equivalence principle to find the field equations of one's choice.\\

In a nut-shell, we assert that Bianchi and Ricci identities are the two fundamental identities to procure scalar-tensor theory of gravity.

\end{document}